# Quantum Mechanics as a Carnapian Language

Iulian D. Toader[1]

## 1. Introduction

It is an experimental fact that quantum mechanics (QM) makes adequate empirical claims, and we can know that QM makes adequate empirical claims, since there is plenty of inductive evidence supporting this fact. But suppose that a rational reconstruction of QM can be given. A rational reconstruction, as Carnap stipulated, would formulate QM in terms of a partially interpreted formal language that includes, along with the logico-mathematical axioms, theoretical sentences and correspondence rules that partially endow the language with empirical meaning by relating theoretical sentences to observation sentences. However, due to Gödel's second incompleteness theorem, without allowing stronger theoretical resources, the consistency of a rational reconstruction of QM is unprovable. But without a proof of consistency, we cannot know that QM makes adequate empirical claims. Thus, our supposition must be denied: *QM cannot be rationally reconstructed*.

One way in which this argument, inspired by Michael Potter's criticism of Carnap's philosophy of arithmetic (Potter 2000), can be resisted is by noting an equivocation between two types of knowledge: *fallible* knowledge, based on inductive evidence, and *infallible* knowledge, based on mathematical proof. There is no reason to dismiss the supposition that QM can be rationally reconstructed, if we can fallibly know that QM makes adequate empirical claims, but at the same time we cannot infallibly know that QM makes such claims. This argument, thus, fails to establish that a rational reconstruction of QM cannot be given.

But are there any good reasons to think that a rational reconstruction of QM *can* be given? And if so, what would such a reconstruction be good for? More specifically, what did Carnap think that a rational reconstruction of QM would be needed for? To address these questions, I will proceed as follows. In section 2, I will describe Carnap's reflections on QM, and I will emphasize his claim that a proper philosophical analysis of QM, including a determination of whether its logic has to be revised, requires a rational reconstruction of the theory. After I recall the successive articulations of the notion of rational reconstruction in the development of Carnap's thought, I will turn in section 3 to a discussion of whether a rational reconstruction of QM is possible, and I will briefly consider two standard criticisms, recently echoed by Richard Healey (2017) and David Wallace (2020). But I will argue that both criticisms can be met, for they both assume that rational reconstruction necessarily requires representationalism. Adopting inferentialism instead would allow for the possibility of formulating QM as a Carnapian language. In section 4, I will briefly revisit Sellars' critique of Carnap, and I will point out that the latter could have indeed adopted inferentialism for non-logical terms, just as he had done for logical terms. I will conclude the paper by clarifying a seemingly worrying circularity problem for my argument, and noting a further line of research, already suggested by Carnap, which is reopened by the possibility claim I defend in this paper.

---

[1] https://sites.google.com/site/ditoader

## 2. Carnap on Quantum Mechanics

Carnap's chapter, "Indeterminism in Quantum Physics", of his 1966 book, starts by presenting some basic principles of QM. Heisenberg's Uncertainty Principle is characterized as "a fundamental law that must hold as long as the laws of quantum theory are maintained in their present form" (Carnap 1966, 284). Carnap duly noted that the limitations entailed by this principle cannot be reduced through any possible improvements of our measuring techniques, since they are not due to the imperfections of our measuring instruments. The mathematical representation of a quantum state by means of a wave function defined on an abstract higher-dimensional space, i.e. on configuration space, is carefully presented. Carnap described the deterministic dynamics of quantum-mechanical systems, governed by the Schrödinger equation, and the probabilistic character of all predictions of the results of any measurements performed on such systems, briefly touching on the QM of macroscopic objects (like satellites) as well. On the basis of his understanding of the basic principles of QM, Carnap offered, throughout the book, his answers to some important philosophical questions, which he obviously thought could already be addressed on that basis.

  For example, on Carnap's view, the necessarily statistical character of quantum-mechanical explanations cannot be regarded as a manifestation of our ignorance, but as an expression of the basic structure of the world, which entails that all physical explanations can only be statistical, under the assumption that all laws of physics reduce to the fundamental principles of QM (such as the Uncertainty Principle) (Carnap 1966, 9). Relatedly, Carnap reported as an "interesting speculation" that QM might indicate that this very structure, and thus presumably its fundamental ontology, including space and time, is all discrete, rather than continuous (Carnap 1966, 89). Furthermore, according to Carnap, QM clearly suggests that there can be no explicit definitions of quantum-theoretical concepts in terms of empirical concepts. But he noted that a more satisfactory answer to the question about the empirical meaning of quantum properties like spin, for instance, would require "an elaborate theory" (Carnap 1966, 221). Carnap also maintained that QM is irrelevant to philosophical debates on the existence of free will. This is because, on his view, indeterminate quantum jumps, though random, cannot play any role in decision making since "it is not likely that these are points at which human decisions are made" (Carnap 1966, 221). But even if they were, that would only make our decisions equally random, and so they would simply not be choices at all, but chances. And even if the range of quantum randomness were much greater than in the actual world, as described by QM, that would only decrease the possibility of free choices.

However, according to Carnap, there are also questions that cannot be addressed properly on the basis of the formulations of QM that Carnap was aware of. These include questions about the logic and language of the theory. He wrote: "The revolutionary nature of the Heisenberg uncertainty principle has led some philosophers and physicists to suggest that certain basic changes be made in the language of physics. [...] The most extreme proposals for such modification concern a change in the form of logic used in physics." (Carnap 1966, 288) Among these proposals, he recalled Birkhoff and von Neumann's (1936) change of the transformation rules, by the replacement of the law of distribution (of conjunction over disjunction) by that of modularity. However, unlike Putnam (1968), Carnap was not ready to take lessons in logic from QM. Rather, he was inclined to think that it was not a change in logic, but the change in the causality structure implied by quantum-mechanical laws (i.e. indeterminism), that made QM revolutionary.

The question whether the logic of physics ought to be revised is not one that could be addressed properly without first presenting "the entire field of physics stated in a systematic form that would include formal logic." (Carnap 1966, 290) Even though he is not fully explicit about it, it is quite clear that Carnap demanded a rational reconstruction of modern physics. But he thought that such a rational reconstruction of modern physics had not been given: "[Its] language is still, except for its mathematical part, largely a natural language; that is, its rules are learned implicitly in practice and seldom formulated explicitly." (Carnap 1966, 291) This tacitly implies that even von Neumann's (1932) and Mackey's (1963) rigorous formulations of QM – widely regarded as mathematical axiomatizations of QM *par excellence* – fail to qualify as rational reconstructions of modern physics, despite their mathematical clarity and their transparent use of the axiomatic method. I will come back to this implication in the next section (for some more details on Carnap's view on QM, see Horvat and Toader 2023).

In any case, note that Carnap did not refer to the lack of systematization of *QM in particular*, but that of *modern physics as a whole*: thus, a rational reconstruction of QM alone (without gravitational and/or spacetime physics) would not provide Carnap with a proper basis for discussing alternative logico-linguistic frameworks. Therefore, it seems fair to say that Carnap's unwillingness to take lessons in logic from modern physics is caused not only by the absence of a fully formalized axiomatization of the latter, but also by its disunity, which is still manifest today in the tensions between quantum and gravitational physics. In any case, his main argument appears to have been as follows: *One can establish that logic needs to be revised in QM only after modern physics as a whole has been rationally reconstructed. But there is no such rational reconstruction, thus the need to revise logic in QM cannot yet be established.* Some may consider this argument, and in particular the very strong conditions on logic revision, rather preposterous. But I will consider the following watered-down version of it: *One can establish that logic needs to be revised in QM only after QM has been rationally reconstructed. But there is no such rational reconstruction, thus the need to revise logic in QM cannot be established.* Of course, we still have no rational reconstruction of QM, in the sense Carnap demanded. But the main question that I am concerned with is whether such a reconstruction would be even possible.

**3. Is a Rational Reconstruction of QM Possible?**

At this point, let me very briefly recall some of the history of the notion of rational reconstruction in Carnap's works (for more historical details, see Beaney 2013). Although Carnap mentioned this notion for the very first time in his 1928 book, *Der logische Aufbau der Welt*, its roots are of course in Hilbert's views on formal axiomatics. In a famous 1917 lecture in Zurich, Hilbert said: "When we assemble the facts of a definite, more-or-less comprehensive field of knowledge, we soon notice that these facts are capable of being ordered. This ordering always comes about with the help of a certain *framework of concepts*... [which] is nothing other than the *theory* of the field of knowledge." (Hilbert 1918, §2) A Hilbertian framework of concepts is, however, according to the early Carnap, not yet a rational reconstruction: "A theory is *axiomatized* when all statements of the theory are arranged in the form of a deductive system whose basis is formed by the axioms, and when all concepts of the theory are arranged in the form of *a constructional system* whose basis is formed by the fundamental concepts." (Carnap 1928, §2) Thus, a theory is rationally reconstructed only if its framework of concepts is (presented as) a constructional system.

      A constructional system is a system of concepts that results from the application of what Susan Stebbing called "directional analysis". Although I do not have the space here to discuss this in detail, here is how she articulated this notion in the case of the system of *Principia Mathematica*: "[This] is not a postulational system. Whitehead and Russell explicitly state that their work 'aims at effecting the greatest possible analysis of the ideas with which it deals.' These 'ideas' are the fundamental concepts of mathematics, which they sought to reduce to their simplest elements. The authors were not concerned to construct a postulational system; they did not seek to obtain *one* out of a set of different postulational systems *any* one of which would yield the required demonstrations

regarding a specific set of mathematical statements. They sought a *single* system such that its primitive concepts and its primitive propositions should yield the whole of mathematics. Thus their system is to be uniquely based on a single set of fundamental concepts and postulates; further, it is to admit only of *one* interpretation. Their aim could be achieved if, *and only if*, the primitive postulates, and primitive concepts, are not merely *postulated*, and taken as *undefined*, but are also *fundamental* in a sense which excludes arbitrary selection. Thus the analysis employed is directional." (Stebbing 1932, 90sq) Arguably, the same holds also for Carnap's constructional system, which he presented in his *Aufbau* (for details, see Toader 2015). If this is correct, then a rational reconstruction is a constructional system in Stebbing's sense, i.e., a system of fundamental, non-arbitrarily selected concepts.

After 1934, however, Carnap's view is that the rational reconstruction of a theory is supposed to use the tools developed in his book *Logische Syntax der Sprache*. More exactly, an empirical theory is rationally reconstructed if presented as a formal language together with a set of semantic or correspondence rules that partially connect this language to its empirical interpretations (Carnap 1939, 60). Rational reconstruction thus requires a reformulation of the empirical theory, a reformulation that rigorously defines a consequence relation, and distinguishes logical and descriptive terms on that basis. As is well known, Carnap's view evolved from this so-called partial interpretation approach, to one that would employ Ramsey sentences and, later, the Hilbert's epsilon operator (see, e.g. Demopoulos 2007, Gratzl and Schiemer 2016).

To come back to my questions then: can there be a rational reconstruction of QM, as a constructional system of fundamental, non-arbitrary concepts or as a partially interpreted formal language? What would recommend any of such reconstructions as a proper condition for an evaluation of logic in QM?

What might recommend rational reconstruction as a proper condition for an evaluation of logic in general, for that matter? In 1943, Carnap maintained that a rational reconstruction of classical logic is needed for its metasemantic evaluation, for the rational reconstruction can explicate properties of classical logic such as whether the meaning of its connectives and quantifiers is fixed. This property is explicated by investigating whether the rational reconstruction of classical logic is complete or *categorical*, i.e., whether it has a unique semantics up to a isomorphism. Thus, on his view, a rational reconstruction is needed for understanding the metasemantics of classical logic. Carnap argued that the rational reconstruction he considered, i.e., a Hilbert-style formal axiomatization of classical logic, is *not* categorical, by constructing what he called non-normal valuations for negation and disjunction, as well as for the quantifiers (Carnap 1943). This raised what came to be called a categoricity problem: same formal rules, different meanings. Importantly, this problem was revealed by investigating a rational reconstruction of classical logic, and led

Carnap to propose a revision of its calculus by allowing arguments that have multiple conclusions.

But what would a rational reconstruction of QM be good for? Would such a rational reconstruction be also required for a metasemantic evaluation of QM? If so, then investigating whether the rational reconstruction is categorical might likewise suggest a revision of the QM formalism, perhaps even a revision of its logical rules.

One way to think about this investigation, I suggest, is as a metatheoretical test for quantum inferentialism. As just mentioned, Carnap applied this test to classical logic, but the same test could of course be applied to theories, such as Peano Arithmetic: Can PA be rationally reconstructed? If it can, then the properties of the reconstruction can explicate some properties of PA. But is the reconstruction categorical? If not, then the arithmetical rules do not fix the meaning of arithmetical terms. Thus, arithmetical inferentialism is false. Suppose we want to run the test on QM: Can QM be rationally reconstructed? If yes, then the properties of the reconstruction can explicate some properties of QM. But is the reconstruction categorical? If not, then the quantum rules do not fix the meaning of QM terms. Thus, quantum inferentialism is false.

The problem with this latter test is, again, that we don't know whether a rational reconstruction of QM is possible. Might standard QM already count as such? Standard QM, i.e., the axiomatization given by von Neumann in 1932, might look like a Hilbertian framework of concepts. But is it a Carnapian constructional system? For one thing, it's not clear that its basic concepts (like the concept of Hilbert space) are fundamental and non-arbitrarily selected, and there is no systematic construction of concepts, as in Carnap's *Aufbau*, either. The standard axioms do not seem to be a result of the kind of directional analysis illustrated in *Principia Mathematica*. Certainly, they do not "admit only of *one* interpretation", as Russell and Whitehead had required for their axiomatization of mathematics. Quite the opposite, in fact.

Recent reaxiomatizations and reconstructions of QM (e.g., Hardy 2001) could be considered results of directional analysis, since they purport to derive standard QM from general information-theoretical principles, considered to be more fundamental and less arbitrary than von Neumann's axioms. But these reaxiomatizations are not formulated as Carnapian languages, and they do not even consider the central task of defining a consequence relation for the language of the reconstructions. Perhaps this is not surprising after all: maybe there is no rational reconstruction of QM because there simply cannot be any such reconstruction, in any of the senses of this notion articulated by Carnap.

This conclusion might be suggested, for instance, by Healey's rejection of the Carnapian reconstructionist project: "For Carnap, the task of the philosopher is to seek clarification of what a scientific theory says by means of a logical reconstruction of the theory within a precisely defined language. ... there is now a consensus among philosophers of science that the labor involved in re-expressing a significant theory in a formal language and then giving its semantics by means of correspondence rules would make this neither a practicable nor a useful technique for revealing its structure and function." (Healey 2017, 123) Interestingly, however, the reported consensus is taken to justify a rejection of rational reconstruction as neither practicable nor useful. But note that Healey does not explicitly maintain it to be an impossible technique. More importantly, note that a rational reconstruction, as in Carnap's partial interpretation approach to scientific theories, is understood to require representationalism, i.e., that the semantics of the formal language of a rational reconstruction is supposed to be given by correspondence rules. This leaves it open that a rational reconstruction were possible, and might be even a practicable and useful technique, if it should not require representationalism, but inferentialism, i.e., that the semantics of the formal language wold be given by inferential rules. Healey, as an inferentialist, would likely embrace this suggestion.

However, others reject Carnap's project precisely because they understand it to require non-representationalism, at least for theoretical terms, which is what Carnap had arguably suggested already in his 1927 paper, *Eigentliche und uneigentliche Begriffe*. Criticizing Healey's own inferentialist approach to standard QM, Wallace writes: "The non-representationalist strategy ... is not new, nor is it specific to quantum theory. It is, rather, the central idea in the logical-positivist and logical-empiricist pictures of science. ... It is almost universally accepted today that these approaches are not viable. But the predominant reason, historically, that they fell from grace was ... the increasingly clear realization – notably ... in the recognized failure of Carnap's project in the *Aufbau* (1928) – that observation is theory-laden." (Wallace 2020, 92) The predominant reason reported here is implied to justify the rejection of rational reconstruction on the basis of a problematic division between observational terms (admitting a representationalist semantics) and theoretical terms (having a non-representationalist semantics). Since no such division is actually defensible, the failure of both Carnap's reconstructionist approach and Healey's inferentialist approach is unavoidable. But this criticism obviously overlooks the very same possibility that I just suggested above, that the rational reconstruction of a theory should dispense with representationalism *not only for its theoretical terms.* If inferentialism were instead globally required, for all the terms in a language, then perhaps the problematic division between observational and theoretical terms would not be enough for the rejection of rational reconstruction.

But would it be possible for the Carnapian interested in rationally reconstructing QM to adopt global inferentialism, both both theoretical and observational terms in a language? To address this question, even if only succinctly, I propose that we revisit Sellars's inferentialist critique of Carnap's view in the *Syntax*. My purpose in the next section is to suggest a positive answer to this question, rather than rejecting that critique and defending Carnap's view against it (for one such defense, see e.g. Carus 2004).

## 4. Carnap's *Syntax* and Sellars' Inferentialist Critique

Carnap's *Syntax* was the inspiration and, at the same time, the target of Sellars' paper "Inference and Meaning" (Sellars 1953): "Sellars identifies his 'material rules of inference' with Carnap's 'P-rules'. ... Carnap's views ... made the scales fall from Sellars's eyes." (Brandom 2015, 43sq) Nonetheless, against Carnap's view that P-rules are just thought-economical devices, "a matter of convention and hence, at most, a question of expedience" (Carnap 1934, 180), Sellars argued that material rules of inference are indispensable, and irreducible to logical rules. In order to explain his argument, consider an example of a logical inference: *If an object is red, then it's colored. This object is red. Thus, it's colored.* And another: *If something is gray, then it's a slithy tove. Findus is gray. Thus, it's a slithy tove.* What is characteristic of such inferences is, of course, that one can assent to their conclusions once one has assented to their premises, even though one may not know what the terms mean. In a slogan, it's the logical form that matters.

Sellars' slogan was, however, different: it's the matter that matters, not the logical form. In the case of material inferences, he contended, unless one assents to their conclusions once one has assented to their premises, one does not know what the terms involved mean. Here is an example of what he considered material inferences: *This object is red. Thus, it's colored.* And another: *Findus is on the mat. Thus, Findus is not on the roof.* Sellars noted that material inferences are expressed by subjunctive conditionals, such as 'If this object were red, then it would be colored.' and 'If Findus were on the mat, Findus would not be on the roof.' But subjunctive conditionals, as Brandom later emphasized, are in fact implicit modal statements: 'Necessarily, if this object is red, then it is colored.' and 'Necessarily, if Findus is on the mat, then Findus is not on the roof.' Due to this implicit modality, Sellars concluded that (the rules of) material inferences are irreducible to (the rules of) logical inferences. This is so, on his view, because one cannot detach the consequent of a subjunctive conditional *only* by affirming its antecedent. For better or worse, Sellars appears to have taken this argument to establish the metasemantic indispensability of (the rules of) material inferences: such inferences must be regarded as relations between meanings, not as relations between extensions of concepts. Expressing material inferences as implicit modal statements makes this point explicit. He also thought that this corrected Carnap's view.

According to Sellars, Carnap had maintained that the material rules are, in principle, reducible to logical rules: "Carnap, however, makes it clear that in his opinion a language containing descriptive terms need not be governed by extra-logical transformation rules. Indeed, he commits himself (p. 180) to the view that for every language with P-rules, a language with L-rules only can be constructed in which everything sayable in the former can be said." (Sellars 1953, 320;

original emphasis) But it is rather hard to endorse this reading of what Carnap actually says on page 180 of his *Syntax*, and one would have to say much more to justify the step from P-rules being a matter of convention and expedience, as Carnap saw them, to P-rules being metasemantically dispensable, in the sense Sellars thought Carnap saw them. On Sellars's reading, Carnap regarded material rules as admissible only on account of "the economy in the number of premises required for inferences" (Sellars 1953, 321). So Sellars considered Carnap's view to be that material rules are inessential to any language, and thus metasemantically inert. What Carnap actually maintained is that, in principle, one could stipulate only logical rules or one can adopt as material rules all sentences that are not logical rules: the choice is based on pragmatic criteria like simplicity and fruitfulness. He never said that adding or dropping P-rules would leave the expressive power of a language unchanged. In any case, Sellars doubted that the view that material rules are metasemantically dispensable is correct: "But might it not be possible for an empiricist to hold that material rules of inference are as essential to meaning as [the logical] rules? ... P-rules are essential to any language which contains non-logical or descriptive terms." (Sellars 1953, 336) Indeed, Sellars believed that his argument from the implicit modal character of material inferences successfully refutes what he took to be Carnap's view.

But Carnap's reflections on P-rules actually raise an important point for Sellars' inferentialism, and furthermore emphasize that the two views are actually compatible. Carnap wrote: "If P-rules are stated, we may frequently be placed in the position of having to alter the language; and if we go so far as to adopt all acknowledged sentences as valid, then we must be continuously expanding it. But there are no fundamental [as opposed to practical] objections to this." (Carnap 1934, 180) Continuous expansion of a language and continuous modification of its semantics, although impractical, may be admissible. But on Sellars' inferentialism, such a modification would be unavoidable. The introduction of new vocabulary, and as a special case the introduction of new concepts in science, always allows us to make novel material inferences. Given the metasemantic work done by such inferences, on Sellars' view, they will not only determine the meaning of the new vocabulary, but will change the meaning of at least some of the old expressions as well. This updating of semantics presupposes, in any case, that material rules are such that novel material inferences can be made when new concepts are introduced in a language.

Moreover, as Carnap rightly noted, rules can also be altered: "No rule of the physical language is definitive; all rules are laid down with the reservation that they may be altered as soon as it seems expedient to do so. This applies not only to the P-rules but also to the L-rules, including those of mathematics. In this respect, there are only differences in degree; certain rules are more difficult to renounce than others." (Carnap 1934, 318) Thus, for Carnap, updating the semantics of a language as a result of changing the rules of inference would be admissible as well, while for

Sellars this updating would be, once again, unavoidable. Since material inferences determine the meaning of empirical terms, every change of rules implies a modification of the meaning of at least some of these terms in the language.

Thus, whereas for Carnap, the continuous expansion of an empirical language, the modification of its rules, and the updating of its semantics, are all admissible, for Sellars they are unavoidable. This suggests that if good reasons had become apparent to him, Carnap would not have dismissed an inferentialist semantics for empirical vocabulary. Carnap adopted inferentialism with respect to logical terms, as can be seen most clearly in his preface to the *Syntax*: "Let any postulates and any rules of inference be chosen arbitrarily; then this choice, whatever it may be, will determine what meaning is to be assigned to the fundamental logical symbols." (Carnap 1934, xv, emphasis removed) But if my observations above are correct, then it seems clear that Carnap could have embraced inferentialism, not only with respect to logical terms, but with respect to non-logical, empirical terms as well, for there was no principled reason for him to resist it.

## 5. Conclusion

My argument, as presented so far, seems to have a circularity problem: for it first stipulates that testing inferentialism requires a rational reconstruction of QM, but then emphasizes that a viable rational reconstruction of QM would better drop representationalism and adopt, instead, global inferentialism. However, there really is no circularity, for the argument actually says that testing inferentialism *for the language of QM* requires a rational reconstruction of QM, but a viable rational reconstruction of QM requires adopting inferentialism *for the language of that reconstruction*. There is no circularity involved in a test of inferentialism for the language of QM that requires adopting inferentialism for the language of the reconstruction.

In 1966, Carnap issued a call "for close cooperation between physicists and logicians -- better still, for the work of younger men who have studied both physics and logic. The application of modern logic and the axiomatic method to physics will, I believe, do much more than just improve communication among physicists and between physicists and other scientists. It will accomplish something of far greater importance: it will make it easier to create new concepts, to formulate fresh assumptions." (Carnap 1966, 291) If the argument in this paper is sound, then it suggests that the Carnapian project of the rational reconstruction of a scientific theory like QM is still alive, despite an apparent consensus to the contrary. A viable rational reconstruction might, as Carnap hoped, lead to the creation of new concepts and new rules for QM. Whether this is something that can indeed be accomplished by rational reconstruction is a question worth investigating in further research.